# Wafer-Scale Lateral Self-assembly of Mosaic Ti$_3$C$_2$T$_x$ (MXene) Monolayer Films

Mehrnaz Mojtabavi[1], Armin VahidMohammadi[2], Davoud Hejazi,[3] Swastik Kar,[3] Sina Shahbazmohamadi[2], Meni Wanunu[3,4*]

[1] Department of Bioengineering, Northeastern University, Boston, MA 02115, USA.

[2] Innovation Partnership Building, UConn TechPark, University of Connecticut, Storrs, CT 06269, USA

[3] Department of Physics, Northeastern University, Boston, MA 02115, USA.

[4] Department of Chemistry and Chemical Biology, Northeastern University, Boston, MA 02115, USA

E-mail: wanunu@neu.edu

## Abstract

Bottom-up assembly of two-dimensional (2D) materials into macroscale morphologies with emergent properties requires control of the material surroundings, so that energetically favorable conditions direct the assembly process. MXenes, a class of recently developed 2D materials, have found new applications in areas such as electrochemical energy storage, nanoscale electronics, sensors, and biosensors. In this report, we present a lateral self-assembly method for wafer-scale deposition of a mosaic-type 2D MXene flake monolayers that spontaneously order at the interface between two immiscible solvents. Facile transfer of this monolayer onto a flat substrate (Si, glass) results in high-coverage (>90%) monolayer films with uniform thickness, homogeneous optical properties, and good electrical conductivity. Multiscale characterization of the resulting films reveals the mosaic structure and sheds light on the electronic properties of the films, which exhibit good conductivity over cm-scale areas.

**Keywords:** Self-assembly, MXene, 2D materials, 2D titanium carbide, monolayer, conductive

**Introduction**

In recent years, various types of two-dimensional (2D) materials have gained much attention due to their unique properties, and have played significant roles in the development of new types of electrical devices[1-3], chemical/biological sensors[4-7], water purification membranes[8-9], energy conversion/storage devices[1, 10-11], and other applications. Despite widespread theoretical and experimental research into the physics and applications of 2D materials, a practical challenge common to all of these materials are related to their handling, where their extremely thin nature and ultrahigh aspect ratios (1,000-10,000) demand methods for manipulation and assembly of these materials during device manufacturing. For example, in applications where a controlled 2D material film is required for harvesting the material's electrical and optical properties over macroscale areas, techniques for large-scale assembly of 2D flakes into uniform mono- and multi-layer films are in need. Efforts to address this ongoing challenge has resulted in the development of several large-scale thin-membrane assembly methods. These methods can be categorized into two groups: 1. assembly of ultrathin films from a 2D material flake dispersion, either directly on a substrate or at an interface (liquid/liquid or liquid/air)[12-13], 2. direct large-area Chemical Vapor Deposition (CVD) of 2D materials on a substrate, followed by use or transfer to another substrate[14-15].

Among 2D materials, graphene oxide (GO) and its reduced counterpart (rGO) have been most widely studied for large-scale monolayer film assembly from their dispersions. Precise methods for casting uniform layers of GO/rGO on substrates include the Langmuir method[16-21], spin-coating[19, 22-23], and layer-by-layer assembly[24-25]. The Langmuir and spin-coating methods result in well-packed films with organized structure and controlled thickness[23] while other methods result in a looser, more disordered structure[12]. For other 2D materials such as graphene, $MoS_2$, and $WS_2$, poor stability in water and other common solvents limits the quality of their assembly into a film from a dispersion phase[26-28]. Their wafer-scale CVD synthesis, however, has been shown to result in more uniform and reproducible films[29-35].

2D transition metal carbides and nitrides (MXenes) as an emerging class of electrically conductive, hydrophilic, and optically active 2D materials have been subjects of numerous studies in recent years[2, 36]. Their synthesis from MAX phases, in which the MXene layers are bound together by group 13 elements such as Al, Si, etc. (A layer in MAX phase structure), is well established and involves selective etching of the A elements and MXene exfoliation in aqueous fluorine-containing acidic solutions[37]. The resulting MXenes have a general formula of $M_{n+1}X_nT_x$ where M is a transition metal, X is carbon/nitrogen, n can be 1, 2, 3, or 4, and $T_x$ represents oxygen, hydroxyl, and fluorine groups present on the MXene basal planes after their exfoliation. Assembly and fabrication of ultrathin (i.e. monolayer) MXene films, however, has so far been challenging and limited the implementation of these materials into devices that require nanometer-thick membranes. Layer-by-layer assembly[38] has shown potential for controlled assembly of MXene monolayer and multilayer films, and other methods have been employed for MXene assembly into thin films, including spin coating[39], spray coating[40-41], and dip coating[42].

Recently, an interfacial assembly method was used for assembling $Ti_3C_2T_x$ MXene layers at an air/water interface, assisted by ethyl acetate[43]. However, in this work the number of layers varied significantly during deposition, possibly due to a strong convection induced by rapid ethyl acetate evaporation. **Table S1** highlights the advantages and disadvantages of large-scale ultrathin film fabrication methods with single-layer precision.

We recently developed a method for lateral self-assembly of MXenes on a liquid-liquid interface to fabricate monolayer to multilayer $Ti_3C_2T_x$ and $Ti_2CT_x$ MXene films on small scale (with tens of microns lateral size) and used them as nanometer-thin freestanding membranes in solid-state nanopore sensing applications[44]. Herein, we have further refined our rapid interfacial self-assembly technique to enable large-scale fabrication of monolayer to few-layer $Ti_3C_2T_x$ films with control over the nominal film thickness, flake density, and lateral size of the self-assembled films. We find that careful control over the MXene suspension concentration affords large-area films with monolayer coverage values that exceed 90%, as characterized by transmission electron microscopy (TEM), scanning electron microscopy (SEM), and atomic force microscopy (AFM) techniques as well as Raman mapping and ellipsometric mapping. Further, successive depositions of monolayer films result in bilayers and trilayers as indicated by optical transmission spectra and electrical sheet resistance measurements. Our lateral self-assembly method results in a mosaic-type high-coverage monolayer MXene films achieved rapidly (~30 min) and without the need for any specialized instrumentation, paving the way for applications that require large-area uniform MXene films.

## Results and Discussion

We synthesized $Ti_3C_2T_x$ MXene from its MAX phase ($Ti_3AlC_2$) according to the MILD synthesis method[45] (see **Materials and Methods** section for details). Previously, we have shown using $Ti_3C_2T_x$ and $Ti_2CT_x$ MXenes that liquid−liquid interfacial self-assembly can be used to generate large-area suspended flake assemblies, which facilitates transfer onto substrates without the need for alignment of flake and substrate feature (e.g., nanoaperture)[44]. In this study, we refined and scaled-up our approach to the wafer scale, as schematically illustrated in **Figure 1a.** After chemical exfoliation of $Ti_3C_2T_x$ and delamination of the monolayer flake dispersion in water, we diluted the dispersion to an optimal concentration controlled by UV-vis absorbance spectrometry using an Eppendorf BioSpectrometer (New York, USA) for monolayer assembly with (**Figure S1**). Then, we added methanol to the MXene dispersion in a 1:8 methanol:water volume ratio (step i). Next, we submerged the substrate of interest in a chloroform bath in a PTFE dish, and drop cast a few droplets (ca. 500 µl) of the prepared mixed dispersion on the chloroform to initiate the self-assembly process (step ii). Immediately following this, flakes begin to coalesce at the chloroform-water interface, forming a large-area two-dimensional film within 5-10 minutes depending on the final film dimensions. Once assembled, as visually seen by no increase in the overall size of the assembly, the film is transferred to the substrate by slowly emerging it from the chloroform bath, followed by drying on a hot plate at 100°C for 30 minutes to complete the transfer process (step iii). Depending on the desired application or characterization method, different substrate types and sizes could be used.

To investigate the properties of the fabricated films and their potential applications, we studied structure, uniformity, fractional area of coverage, optical and electrical properties of the fabricated films. **Figure 1b** shows an AFM image of the transferred monolayer $Ti_3C_2T_x$ film on a $SiN_x$ substrate (see also **Figure S2 (a-i)**). These images confirm the mosaic structure and effective monolayer nature of the films with well-packed and edge-to-edge arrangements and partial overlaps at the edges. The corresponding line height profiles of monolayer flakes are shown in **Figure 1b** and a monolayer $Ti_3C_2T_x$ thickness is measured around 1.8 nm (see also **Figure S2g**). Theoretically, the thickness of a $Ti_3C_2T_x$ flake is 0.98 nm; however, height artifacts in AFM are common for hydrophilic materials due to presence of water beneath/above the flake, presence of functional groups on the MXene basal plane, and interactions of the bound surface groups of $Ti_3C_2T_x$ flakes with the substrate, all resulting in a typical AFM-based thickness of 1.5 - 2 nm for $Ti_3C_2T_x$ monolayer flakes[46-47], which our measurements agree with. To further corroborate the film thickness, we performed ellipsometric measurements. In **Figure 1c, d,** we present an optical image and an ellipsometric thickness map of a 3" Si wafer onto which we deposited a monolayer of $Ti_3C_2T_x$. The optical image shows a partially covered wafer before the complete evaporation of water. Around the edges of the wafer there is no coverage, which provides an optical contrast as indicated by arrows in the inset image (blue = MXene monolayer, red = bare Si). To assess the thickness variation across the wafer, we mapped the ellipsometric quantities Ψ and Δ as a function of wavelength at 81 points along the wafer and produced a color map using Voronoi interpolation that represents the MXene layer thickness. The dashed yellow circle shows the area (4 cm in diameter) within which ellipsometric mapping was carried out, and the yellow oval shape on the image roughly represents the size/shape of the ellipsometric incident light beam (~1.2 mm beam diameter). Using these data and published optical data for n, k for a $Ti_3C_2T_x$ film for various wavelengths (see **Figure S2**)[39], we constructed a 3-layer model of our sample (Si, $SiO_2$, and MXene layer), and fit our mapping data to obtain a thickness map (see **Materials and Methods** section for more details). The areas covered with film show thickness values in the range of 0.9 - 1.5 nm, which we speculate corresponds to monolayer $Ti_3C_2T_x$ film along with some areas containing partial bilayer due to laterally overlapping flakes in the assembly.

For a deeper insight into the quality of the ellipsometric data, we show in **Figure 2a** violin plots of Ψ and Δ values measured at different points on the mapping experiment in **Figure 1d** and compare these with values for a bare Si substrate. The difference between Ψ and Δ values for these two substrates confirms the presence of the film over the entire area of the wafer. For each wavelength measured, we find a normally-distributed variation in Ψ and Δ values for the MXene monolayer film. An analogous map of the measured $SiO_2$ thickness is shown in **Figure S5**. Moreover, the outlier points of the violin plots fit the data for bare Si substrate, which identifies a few uncovered regions on the wafer (**Figure 1c**, red arrow, and **Figure 1d**, red region).

Another method to assess the coverage and uniformity of $Ti_3C_2T_x$ films on a substrate is Raman spectroscopy. Sarycheva *et al*[48] recently showed that the Raman spectrum of a single $Ti_3C_2T_x$ flake has two characteristic peaks, and further, observed an increase in peak intensities upon multilayer stacking of these flakes. Unlike transition metal-dichalcogenides (TMDs), multilayer $Ti_3C_2T_x$ flakes didn't show any peak shifts, which relates to the substantial interlayer gap that exists between MXene sheets (~5-6Å)[2]. **Figure 2b** shows a Raman spectrum of monolayer $Ti_3C_2T_x$ film on a Si/$SiO_2$ substrate, which shows two characteristic $Ti_3C_2T_x$ peaks centered at ~300 cm$^{-1}$ and ~975 cm$^{-1}$. The strong peak at 521 cm$^{-1}$ corresponds to the Si substrate. Therefore, to provide some assessment of coverage we obtained Raman spectra from a 68.5 μm x 44.5 μm area and mapped the peak height intensity of MXene peak at ~300 cm$^{-1}$ to the Si peak. While we cannot derive thickness values based on these spectra, we find that MXene peaks are everywhere in the region, fluctuating in relative intensity by <25%. Three spectra (a, b, and c) from three different areas (green, yellow, and orange) on the map in **Figure 2b** show the mean and extremities of our measurements.

Next, to gain a deeper insight into the packing density and fractional area coverage of the films we used AFM, SEM, and TEM characterization. To allow us to inspect the resulting MXene films using electron microscopy, we lifted MXene monolayers onto 5x5 mm$^2$ Si chips hosting freestanding 50-nm-thick silicon nitride ($SiN_x$) membranes at their centers. **Figure S6** show SEM images of monolayer $Ti_3C_2T_x$ film transferred on the Si chips. The $SiN_x$ membranes appear as dark areas beneath the $Ti_3C_2T_x$ film because electrons are only weakly scattered from the ultrathin membrane. We used ImageJ software[49] to calculate the fractional area covered with monolayer $Ti_3C_2T_x$ films based on finding the optimum threshold intensity that differentiates the substrate from the monolayer and quantifying the substrate area. In **Figure 3a-c,** we show AFM, SEM, and TEM images of MXene monolayer films, respectively, along with corresponding post-analysis images below after thresholding, which were used in our coverage analysis. For the AFM measurements we used a SiNx substrate, whereas for SEM and TEM measurements we used 50-nm-thick freestanding $SiN_x$ membranes supported by a Si chip. For these images we find MXene coverage values of 93%, 91%, and 84.1%, respectively, with similar coverage values obtained in other areas we inspected (see **Figures S7-S8** for other images). We conclude that our lateral assembly method produces a mosaic structure of closely-packed flakes that cover >90% of the substrate following transfer.

So far, we have shown that the lateral self-assembly method is capable of producing ultrathin, uniform monolayer and packed $Ti_3C_2T_x$ films. To investigate sequential assembly of bilayers and trilayers, we repeated our transfer method by lifting a monolayer film onto a substrate, drying the substrate, re-immersing the substrate in chloroform, repeating lateral self-assembly of MXene flakes at the interface, and emerging the substrate from the chloroform phase. We examined the resulting multilayer films using optical spectroscopy and electrical conductivity measurements. **Figure 4a** shows the UV-vis absorbance spectra of monolayer, bilayer, and trilayer $Ti_3C_2T_x$ films on a glass substrate. As seen in the spectra, absorbance increases with increasing number of layers. Inset shows the absorbance for monolayer, bilayer, and trilayer films at 550 nm, showing that although the increase is not linear, it is in qualitative agreement with previous studies[43].

Due to the inherent in-plane conductivity of $Ti_3C_2T_x$ flakes, we examined how the number of MXene layers affects the electrical properties of the films. We used the van der Pauw method (vdP) to measure the sheet resistance of the films[50] using a home-made apparatus and a Keithley 2401 source meter (Tektronix, Inc., USA). **Figure 4b** shows the schematic of the apparatus, which consists of a square grid of four gold electrodes separated by 10 mm. Four wires connect to each electrode at points 1, 2, 3, and 4 and attach to the Keithley source meter (see **Figure S9** for an image of the experimental set-up). **Figure 4b** also shows a schematic top-view image of the electrodes and depicts the four edges of the film, two horizontal (H1 and H2), and two vertical (V1 and V2), where conductivity measurements are carried out. First, current is sourced along each edge (horizontal and vertical) and voltage is measured along the opposite edge. This is repeated for all combinations, which results in four independent measurements. Current vs. voltage (IV) curves of these measurements for a monolayer $Ti_3C_2T_x$ film transferred onto a glass substrate are shown in **Figure 4b.** Due to the uniformity of the film within this 10 mm x 10 mm area, the four IV curves are identical. **Figure 4c** shows one representative IV curve for each monolayer, bilayer, and trilayer $Ti_3C_2T_x$ films measured after the sequential transfer of each layer. From the resistance values obtained by IV measurements, sheet resistance values can be calculated (see **Materials and Methods** section for details). **Figure 4d** shows the sheet resistance values for two samples of $Ti_3C_2T_x$ films on a glass substrate with a thickness of 1, 2, and 3 layers. As expected, as the thickness of the film increases, the sheet resistance decreases. In general, our films show higher sheet resistance (9-11 kΩ/sq) compared to the recent study by Yun *et al*[43], which showed resistance of 1.5 kΩ/sq for a monolayer film. We attribute this discrepancy to the mosaic morphology of our films, in which flake-to-flake contact is edge-to-edge with minimal overlapping. Despite the higher resistance we find that our films exhibit ohmic response and that resistance drops with increasing number of layers.

Finally, we have compared the advantages and disadvantages of the various 2D material coating methods in **Table S1**. Compared to other methods, one advantage of our wafer-scale lateral self-assembly is the resulting uniform film structure across the entire film area, which is unlike the spin-coating method that results in a compact monolayer film only at the center of the substrate. Moreover, compared to the Langmuir transfer method, our assembly method is much faster (under 30 minutes for film formation on a 3" wafer scale), and further, does not require any special apparatus apart from a PTFE or glass dish. It is worthwhile to mention that, as discussed in our previous study[44], the driving force for film formation at the liquid-liquid interface is a combination of repulsion force, capillary force, and surface tension, which results in a delicate organization of the resulting mosaic film structure. Our lateral assembly method thus results in smoother and more organized films than those produced by spontaneous absorption of flakes from solution, for example.

**Conclusion**

In summary, we have shown here that lateral self-assembly of MXene films on a liquid-liquid interface produces uniform mosaic-type monolayers that span over large areas (we have coated up to 3" substrates). We have studied the film morphology of our coated substrates at various scales using a combination of tools that include AFM, SEM, and TEM imaging, as well as ellipsometric and Raman mapping. Moreover, the sequential transfer of single-layer films on a substrate enables the fabrication of multilayer films without peeling off the preceding layers. This was confirmed by the change in sheet resistance and optical properties of the films upon sequential deposition of multilayer films. While we have focused here on large-scale self-assembly of $Ti_3C_2T_x$ flakes, our method should be compatible with a variety of other members of the MXene family. Considering the 30+ different MXene compositions experimentally synthesized so far, our method offers an attractive approach for producing large-area uniform films that may find use in applications that require hydrophilic and electrically conductive films such as transparent electrodes for bioelectronics, illuminated displays, and other devices.

**Materials and Methods:**

**MXenes synthesis.** $Ti_3C_2T_x$ MXene flakes with large lateral sizes were synthesized according to previous reports in the literature[46]. Initially, for every 1 g of the MAX phase, the mixed salt-acid etching solution was prepared by adding 1 g of LiF powder (98.5%, Alfa Aesar) to 20 mL of 6 M HCl solution (ACS grade, BDH) followed by stirring for 15 min to completely dissolve the LiF powder in the acid solution. The etching process was started by slowly adding 1 g of MAX phase powder ($Ti_3AlC_2$ synthesized according to previous work[51] ) to the etching solution. The etching container was placed into an ice bath during the addition of MAX phase to avoid excessive heat generation due to the exothermic nature of the reaction. The etching solution was stirred at 550 rpm (by using a PTFE-coated magnetic bar) continued for 24 h at 35 °C. After 24 h, the solution containing etched MXene multilayers was divided into four different centrifuge vials, and 45 mL DI water was added to them to start the washing process. The solutions were then shaken by hand for 1 min and centrifuged at 3,500 rpm (Eppendorf 5810R) for 3 min, after which the supernatant was poured out. The washing process was continued for several times, each time adding 45 mL DI water followed by manual shaking of the solutions for 2 min, and then centrifuging them at 3,500 rpm for 3 min until a dark green supernatant was observed (pH > 4.5).The supernatant after this stage is called delaminated $Ti_3C_2T_x$ dispersion. The initial delaminated solution (the first supernatant after washing step was complete) was poured out and DI water was added to the sediments. The solutions were shaken for another 2 min and this time centrifuged at 3500 rpm for 1 h to collect the large flake size MXene solutions (pH ~5).

**Wafer-scale Lateral Self-assembly of MXenes and Transfer on Substrate.** Monolayer MXene film formation was carried out by optimizing our previously reported lateral interfacial self-assembly method[44]. MXene dispersion was first diluted to the desired concentration and then mixed with methanol to a 1:8 volume ratio. The substrate of desire was submerged in a chloroform bath and dispersion was added to the chloroform until the droplet size enlarged to the size of the substrate. After flakes were self-assembled on the droplet/chloroform interface, excess chloroform was removed from the bath and the film level lowered down to reach the substrate. Afterward, the substrate was lifted and baked at 100°C for ~30 min to dry. The substrates were kept in a vacuum desiccator until use.

**Imaging Techniques.** For SEM and TEM imaging, monolayer $Ti_3C_2T_x$ films were transferred onto Si chips with fabricated 50 mm x 50 mm 50-nm-thick $SiN_x$ membranes at their center. For AFM imaging, a film was transferred onto $SiN_x$ substrates. High-resolution TEM imaging was done using a JEOL 2010FEG operating in bright-field mode at 200 kV. Scanning electron microscopy (SEM) was done using JSM-IT200 at 3.0 kV accelerating voltage. AFM measurements were carried out by the FastScan AFM instrument (Bruker Instruments, Billerica, MA) using ScanAsyst cantilevers (Bruker Instruments) performing at the FastScan device's ScanAsyst Mode.

**Ellipsometry Spectroscopy and Mapping:** Ellipsometry measurements were carried out using HORIBA UVISEL 2 Ellipsometer (HORIBA Scientific, USA) on monolayer $Ti_3C_2T_x$ films deposited on Si substrate. Real (n) and imaginary (k) refractive indices previously developed by Dillon *et al*[39] were used to make a reference file for $Ti_3C_2T_x$ using DeltaPsi2 software. Using reference files for modeling, the optical constants are fixed for fitting, and only the thickness changes. We produced the model using DeltaPsi2 software based on the layered structure, consisting of a Si layer, a $SiO_2$ layer (due to the presence of native oxides on Si surface), and a $Ti_3C_2T_x$ layer (**Figure S4 inset**). **Figure S4** shows how well our model fits the measured Ψ and Δ values obtained from a monolayer $Ti_3C_2T_x$ film. Based on these parameters, the thickness of the film is calculated to be 0.95 Å which matches the theoretical thickness of a single $Ti_3C_2T_x$ layer. For ellipsometry mapping, the DeltaPsi2 software mapping recipe feature was used to design automated measurements at a spectrum range of 300 to 800 nm at the 70-degree incidence angle on a pre-designed circular grid with a 40 mm diameter and 81 points. The measurement results (Ψ and Δ values) at each point were fitted to the layered model consisting of $Si/SiO_2/Ti_3C_2T_x$ to calculate the thickness. Using X, Y coordinates of grid points, and thickness values, the Voronoi interpolation method was adopted, using Igor Pro software, to map the measurement results to the whole wafer area.

**Raman Spectroscopy and Mapping:** The Raman measurements were carried out using a Thermo Scientific DXR2xi Raman Imaging Microscope on monolayer $Ti_3C_2T_x$ films deposited on 300-nm-thick $SiO_2$ substrate. The excitation source was a 455 nm laser with 0.5 μm spot size. The laser power was kept at 2 mW with a 1 sec exposure time for 10 scans per position. Mapping was done on a 68.5 μm x 44.5 μm area using OminiCX Software.

**UV-vis Absorbance Measurement:** UV–vis absorbance measurements were carried out using a PerkinElmer Lambda 35 UV− vis−NIR spectrophotometer on monolayer, bilayer, and trilayer $Ti_3C_2T_x$ films deposited on glass substrates in the range of 330 to 1100 nm. The effect of the glass substrates was removed by placing an identical clean glass slide in the reference beam position of the spectrophotometer, and hence, the absorbance spectra (shown in **Figure 4a**) are representative for the $Ti_3C_2T_x$ films without any contribution from the substrate.

**Conductivity Measurement:** The conductivity measurements were carried out using Keithley 2401 source meter and a home-made apparatus on monolayer, bilayer, and trilayer $Ti_3C_2T_x$ films deposited on glass substrates. The apparatus consists of four gold electrodes located at four corners of a square with a 10 mm length (shown in **Figure 3b**). To calculate sheet resistance, two sets of measurements were carried out. The first set is to measure the voltage drop across each horizontal line of the square (H1 or H2) by the sourcing current across the opposite edge, followed by calculating the resistance from the slope of the I-V curves ($R_{H1}$, $R_{H2}$). The second set is to measure the voltage drop across each vertical line of the square (V1 or V2) by sourcing current across the opposite edge ($R_{V1}$, $R_{V2}$). After the four measurement sweeps were taken, the sheet resistance was calculated as follows:

$$R_{Horizontal} = \frac{RH1+RH2}{2}$$

$$R_{Vertical} = \frac{RV1+RV2}{2}$$

$$e^{\frac{-\pi R Horizontal}{Rs}} + e^{\frac{-\pi R Vertical}{Rs}} = 1$$ , R$_s$= sheet resistance

**Notes**

The authors declare no competing financial interest.

**Acknowledgment**

We acknowledge the National Science Foundation (DMR 1710211) for funding this work. The authors thank Prof. Yuri Gogotsi and Dr. Aaron Fafarman for providing us the values for modeling ellipsometric measurement. The authors acknowledge Dr. Tzahi Cohen-Karni and Raghav Garg (Carnegie-Mellon University) for loaning us the vdP apparatus. The authors thank Dr. Joshua Gallaway and Matthew Kim for use of their Raman spectroscopy and mapping tool. We thank Dr. David Hoogerheide for providing us with the 3" Si wafer for ellipsometric measurements.

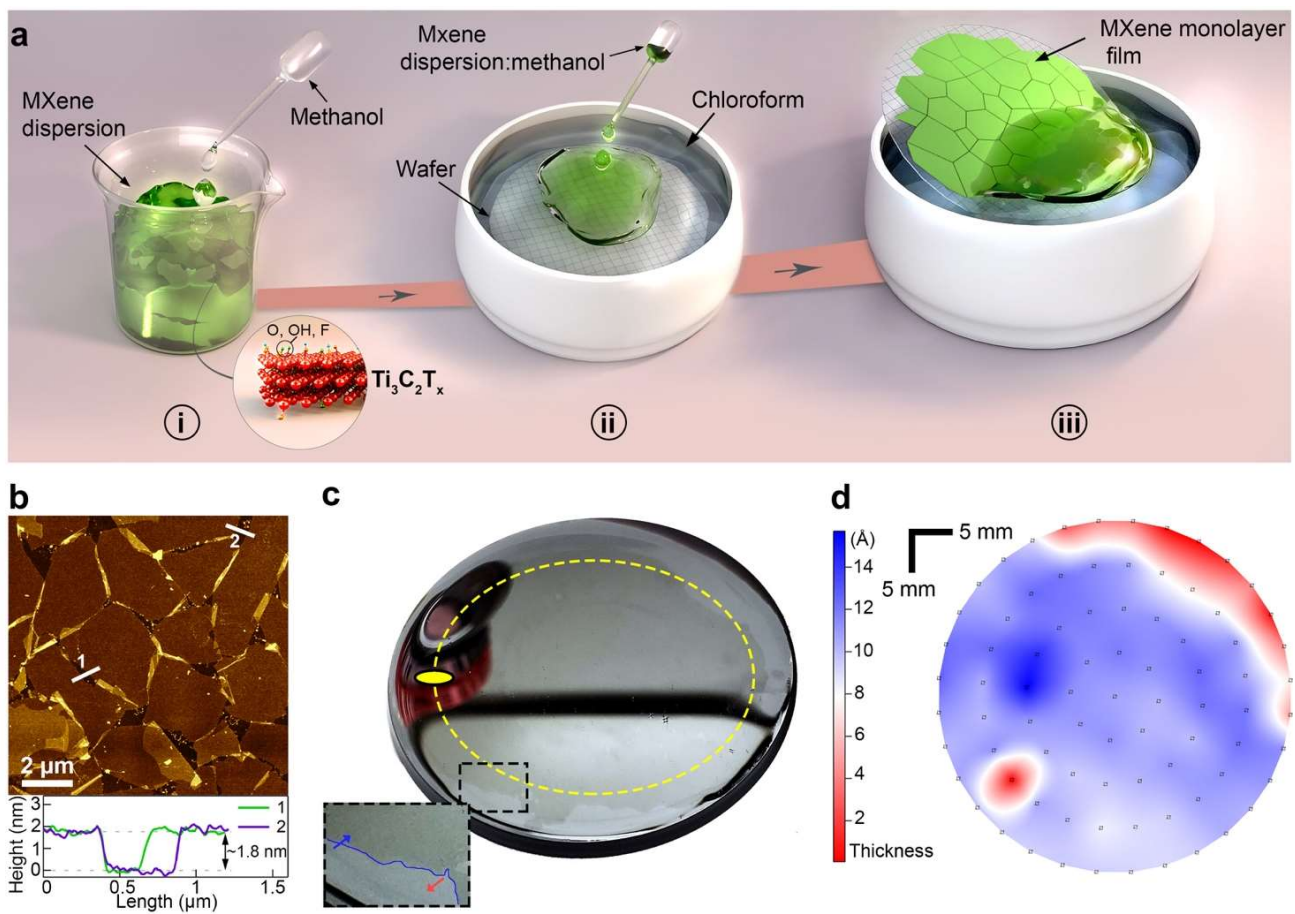

**Figure 1. Lateral self-assembly of monolayer MXenes on a liquid-liquid interface and transfer on a substrate**. (a) Schematic illustration of self-assembly and transfer steps of monolayer $Ti_3C_2T_x$ film. The atomic structure of $Ti_3C_2T_x$ MXene is shown with $O^-$, $OH^-$, and $F^-$ as functional groups on the surface. (b) AFM image of the transferred $Ti_3C_2T_x$ monolayer film on a $SiN_x$ substrate. Corresponding line profiles at the bottom of the image show thickness of ~ 1.8 nm for a monolayer film. (c) Image of the Si wafer with transferred monolayer $Ti_3C_2T_x$ film on top used for ellipsometric measurement. The yellow dashed circle shows the area (with 4 cm diameter) within which ellipsometric mapping was carried out. The yellow oval shape shows the approximate shape and size of the incident beam. The magnified image of the black dashed square shows the edges of the $Ti_3C_2T_x$ film on the Si substrate. The blue arrow shows the film and the red arrow shows the bare substrate. (d) Ellipsometric mapping of $Ti_3C_2T_x$ film thickness on a circular area with a diameter of 4 cm.

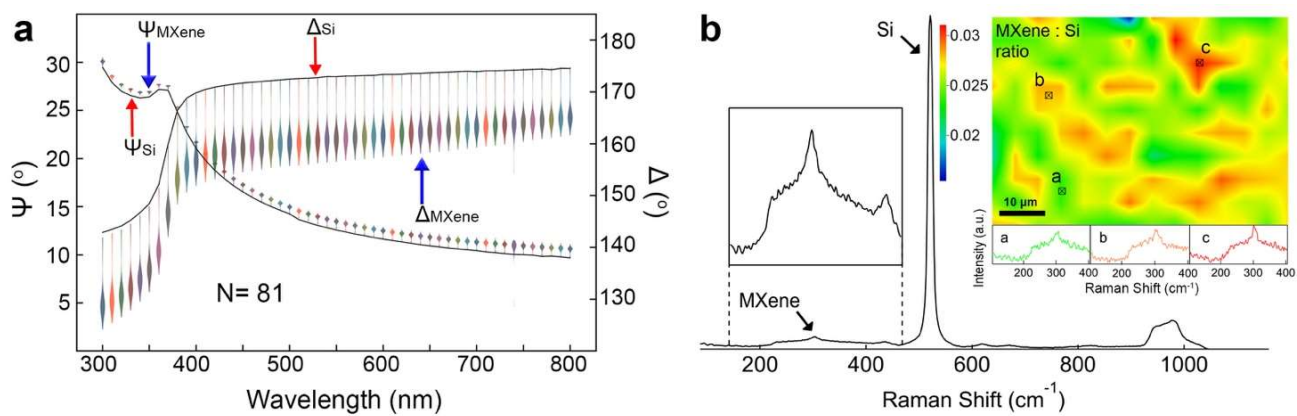

**Figure 2. Characterization of monolayer Ti$_3$C$_2$T$_x$ film uniformity**. (a) Violin plots of obtained Ψ and Δ values from 81 points in a circular area of a transferred Ti$_3$C$_2$T$_x$ film (shown with blue arrows). The red arrows show the Ψ and Δ values obtained from the bare Si wafer. (b) Raman spectrum obtained from a monolayer Ti$_3$C$_2$T$_x$ film on Si/SiO$_2$ substrate. The strong peak of Si at 521 cm$^{-1}$ originates from the substrate. Inset shows the Raman mapping of peak height ratio of MXene: Si for a 68.5 μm x 44.5 μm are. The corresponding Raman spectra of points a, b, and c on the map are shown at the bottom of the image.

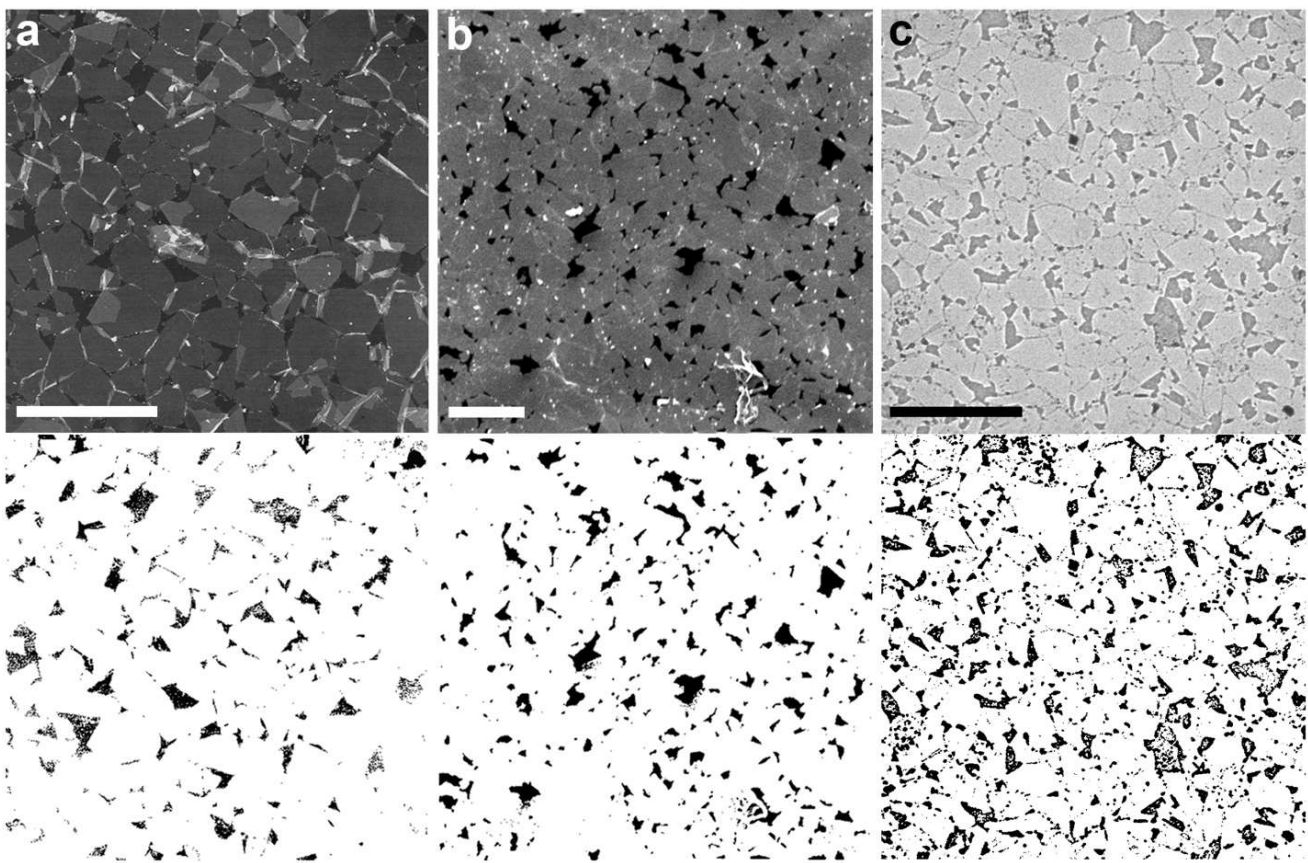

**Figure 3. Characterization of fractional coverage of monolayer $Ti_3C_2T_x$ films**. (a) Top: AFM image of the transferred monolayer $Ti_3C_2T_x$ films on a $SiN_x$ substrate: Bottom: Analyzed image using ImageJ software showing 92.8% coverage of substrate by the film. (b) Top: SEM image of a transferred monolayer $Ti_3C_2T_x$ film on 50-nm-thick freestanding SiNx membrane: Bottom: Analyzed image using ImageJ software showing 91.8% coverage of substrate by the film. (c) Top: Bright-field TEM image of the transferred monolayer $Ti_3C_2T_x$ films on a ~50 nm thick SiNx membrane: Bottom: Analyzed image using ImageJ software showing 84.1% coverage of substrate by the film. All scale bars = 10 μm, and black areas represent uncovered areas in all bottom images.

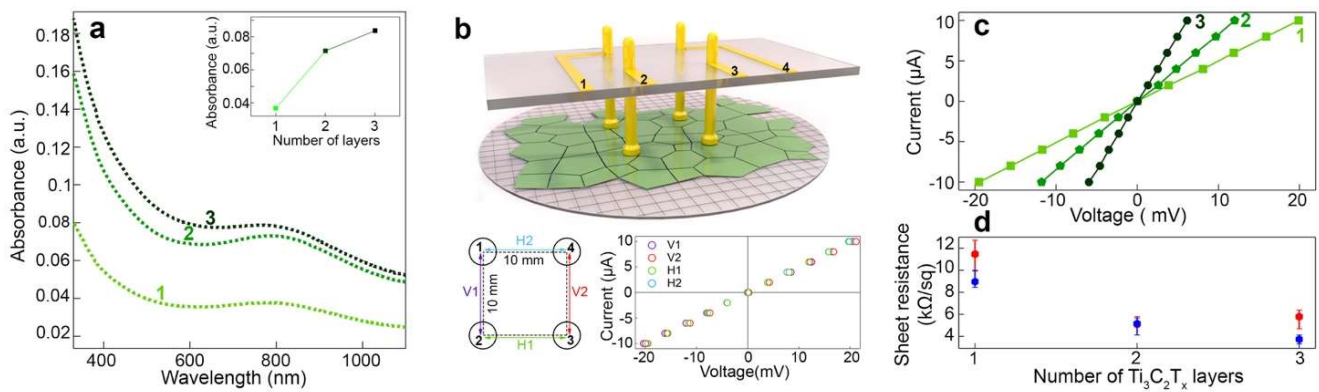

**Figure 4. Optical and electrical characterization of monolayer and multilayer Ti$_3$C$_2$T$_x$ films**. (a) UV-vis absorbance spectra of monolayer, bilayer, and trilayer Ti$_3$C$_2$T$_x$ films. Inset shows the absorbance of the films at 550 nm. (b) Schematic of vdP apparatus used for sheet resistance measurement of the films. Bottom: Top view of the apparatus showing the location of corresponding electrodes from 1-4 and edges of the square area where electrical measurements were carried out (H1, H2, V1, and V2). Corresponding IV curves of current sweep across each edge is shown on the right. (c) Representative IV curves of monolayer, bilayer, and trilayer Ti$_3$C$_2$T$_x$ films. (d) Sheet resistance of Ti$_3$C$_2$T$_x$ films versus the number of Ti$_3$C$_2$T$_x$ layers. Markers show the average of each sheet resistance values and the lines show the range of the values for six measurements on the sample.

# Supporting Information

## Wafer-Scale Lateral Self-assembly of Mosaic Ti$_3$C$_2$T$_x$ (MXene) Monolayer Films

Mehrnaz Mojtabavi[1], Armin VahidMohammadi[2], Davoud Hejazi,[3] Swastik Kar,[3] Sina Shahbazmohammadi[2], Meni Wanunu[3,4]*

[1] Department of Bioengineering, Northeastern University, Boston, MA 02115, USA.
[2] Innovation Partnership Building, UConn TechPark, University of Connecticut, Storrs, CT 06269, USA.
[3] Department of Physics, Northeastern University, Boston, MA 02115, USA.
[4] Department of Chemistry and Chemical Biology, Northeastern University, Boston, MA 02115, USA.

E-mail: wanunu@neu.edu

## Table of Contents

**Table S1**. Comparison between methods of large-scale film fabrication of 2D materials with single-layer precision

| Method of film fabrication | 2D Materials used | Advantages | Disadvantages |
|---|---|---|---|
| **Spin Coating** | Graphene oxide[1-3] | Short fabrication time<br><br>Directly on a substrate, no further transfer is needed | Non-uniform monolayer coverage vs. radial distance from center of substrate. |
| **Langmuir-Blodgett (LB)** | Graphene oxide[2, 4, 5, 6-8]<br>Graphene[9]<br>$Ti_2CT_x$[10] | Well-packed and structurally organized film<br><br>Freedom in lateral size variation | Need for specialized apparatus and skill<br><br>Long fabrication time<br><br>Requires further transfer onto a substrate |
| **Layer-by-layer Assembly** | Reduced graphene oxide[11-12]<br><br>$Ti_3C_2T_x$[13] | No need for specialized apparatus<br><br>Directly on a substrate, no further transfer is needed | Spontaneous absorption (no lateral force)<br><br>Relatively open, loose and disordered film |
| **Interfacial Assembly (liquid-liquid and liquid-air)** | Graphene[14]<br>$MoS_2$ and $WS_2$[15] | No need for specialized apparatus<br><br>Freedom in lateral size variation | Requires further transfer onto a substrate<br><br>low dispersibility of graphene, $MoS_2$, and $WS_2$ in water and common solvents |
| **Interfacial Assembly (liquid-liquid and liquid-air)** | $Ti_3C_2T_x$[16]<br>$Ti_3C_2T_x$ (this work) | No need for specialized apparatus<br><br>Freedom in lateral size variation<br><br>Well-packed and structurally organized film<br><br>Dispersibility of $Ti_3C_2Tx$ in water and common solvents | Requires further transfer onto a substrate |

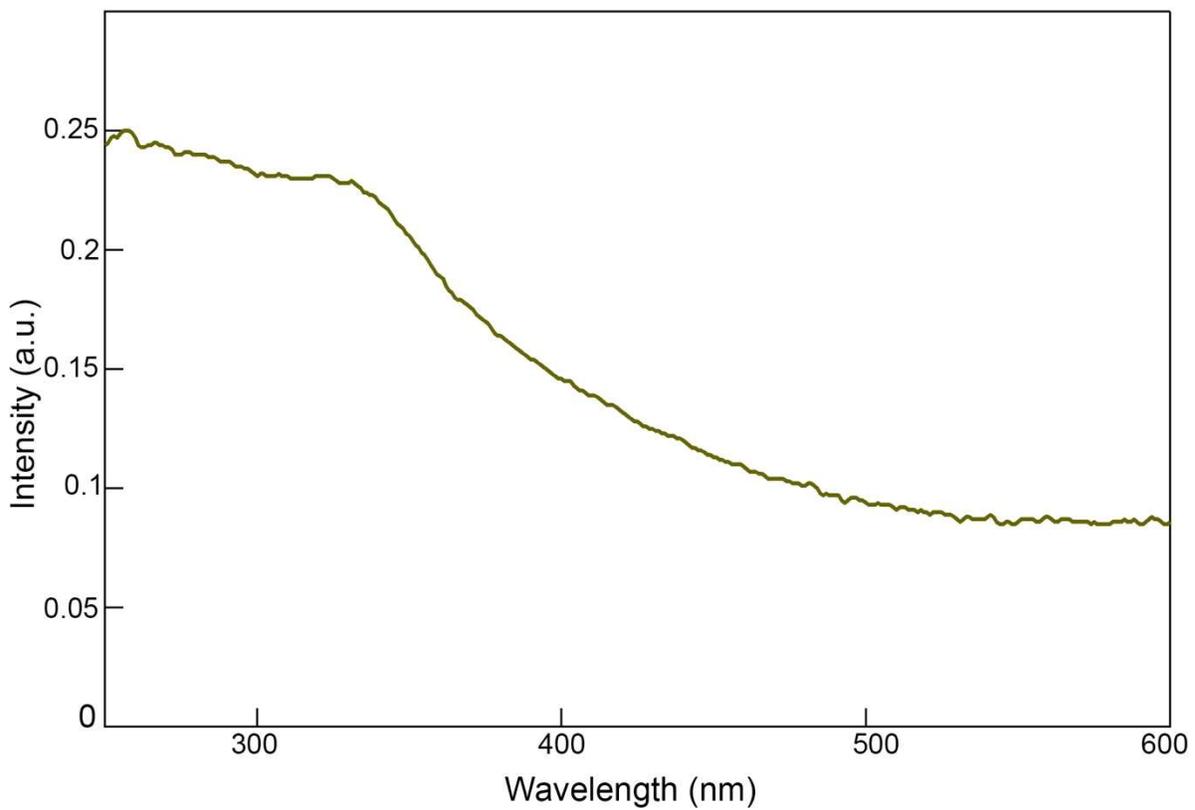

**Figure S1.** UV-vis absorbance of diluted $Ti_3C_2T_x$ dispersion optimized for assembly of monolayer flakes at the liquid-liquid interface.

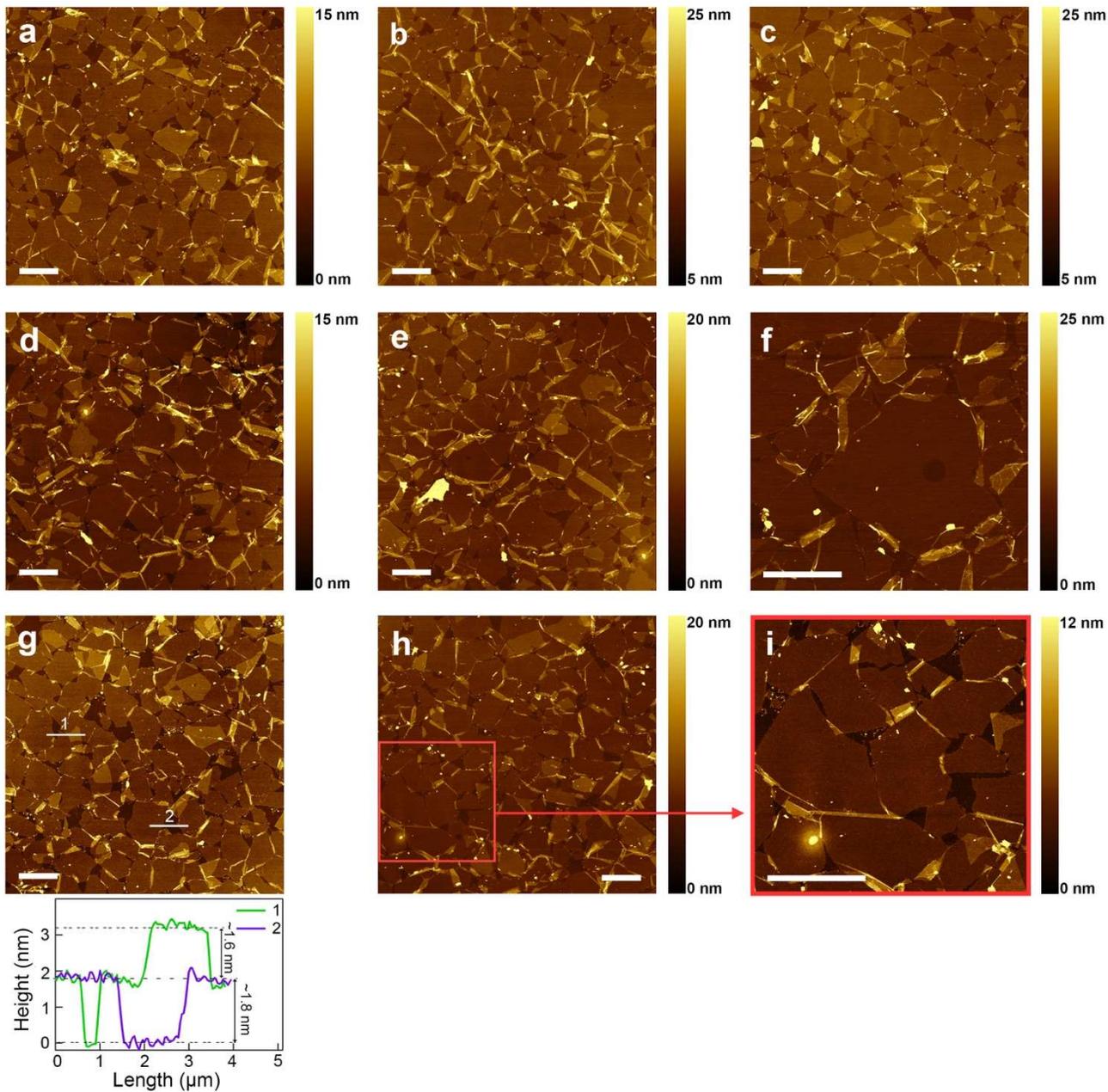

**Figure S2.** (a-i) AFM images of the transferred monolayer $Ti_3C_2T_x$ film on a $SiN_x$ substrate with the corresponding Z scales. Corresponding line profiles of (g) shows a thickness of ~ 1.8 nm for a monolayer film and 1.6 nm at the overlap. (All scale bar = 4 μm)

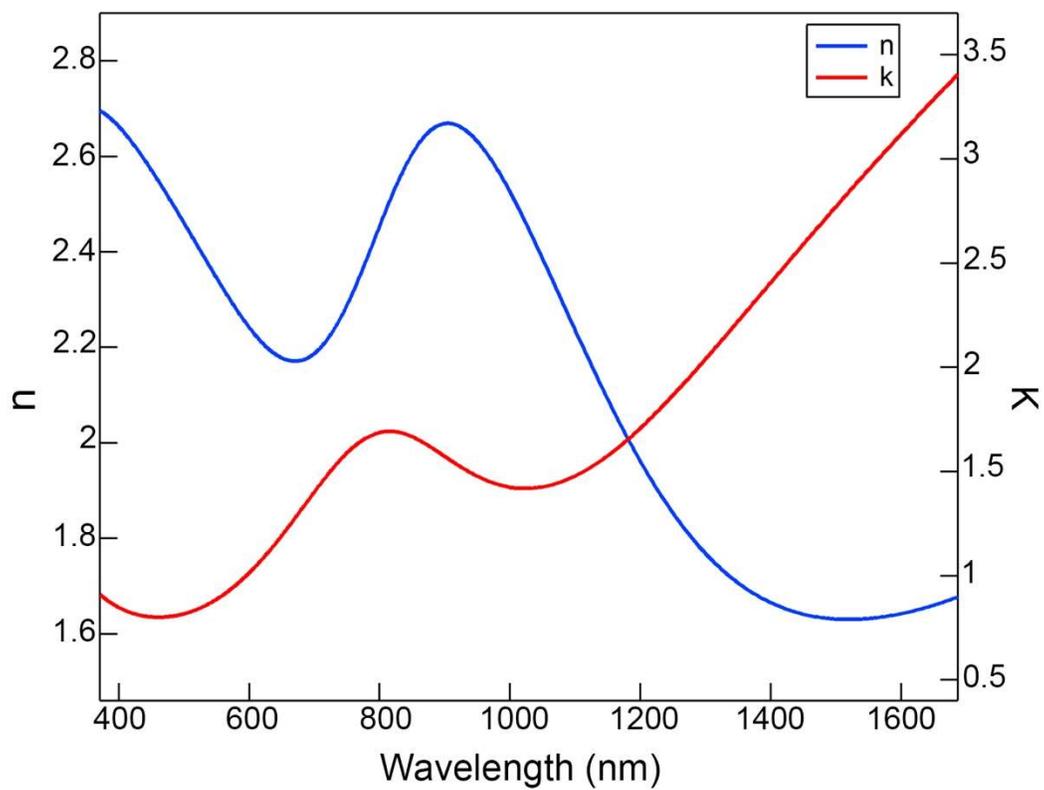

**Figure S3.** Real (n) and imaginary (k) refractive indices of a $Ti_3C_2T_x$ film.

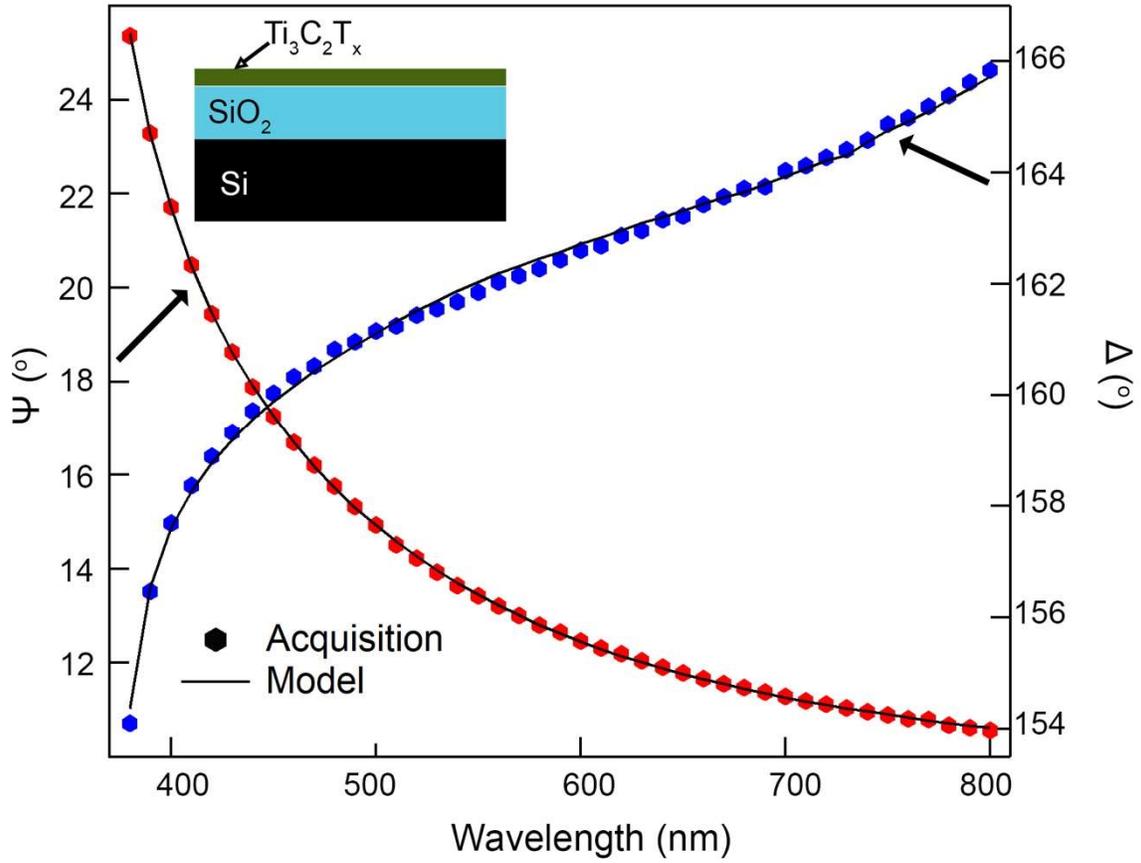

**Figure S4.** Ψ and Δ values obtained from a monolayer Ti$_3$C$_2$T$_x$ film on a silicon substrate and Ψ and Δ values obtained from the model based on the layered structure (Si/SiO$_2$/Ti$_3$C$_2$T$_x$) shown in the inset.

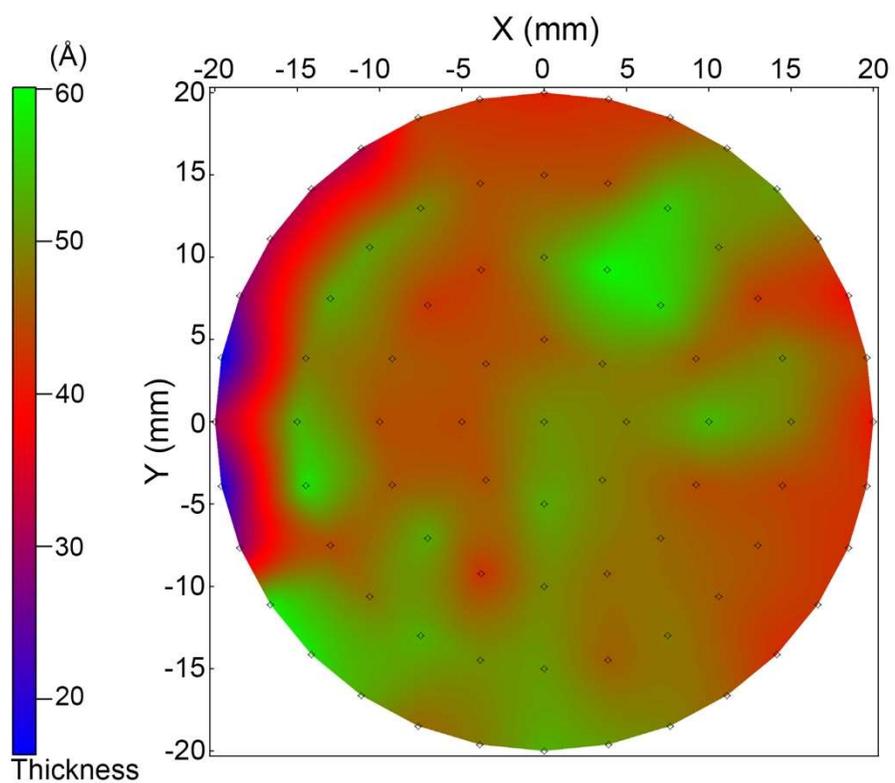

**Figure S5.** Ellipsometric mapping of the SiO$_2$ layer thickness, beneath the MXene layer, over a circular area of a Si wafer with a diameter of 4 cm.

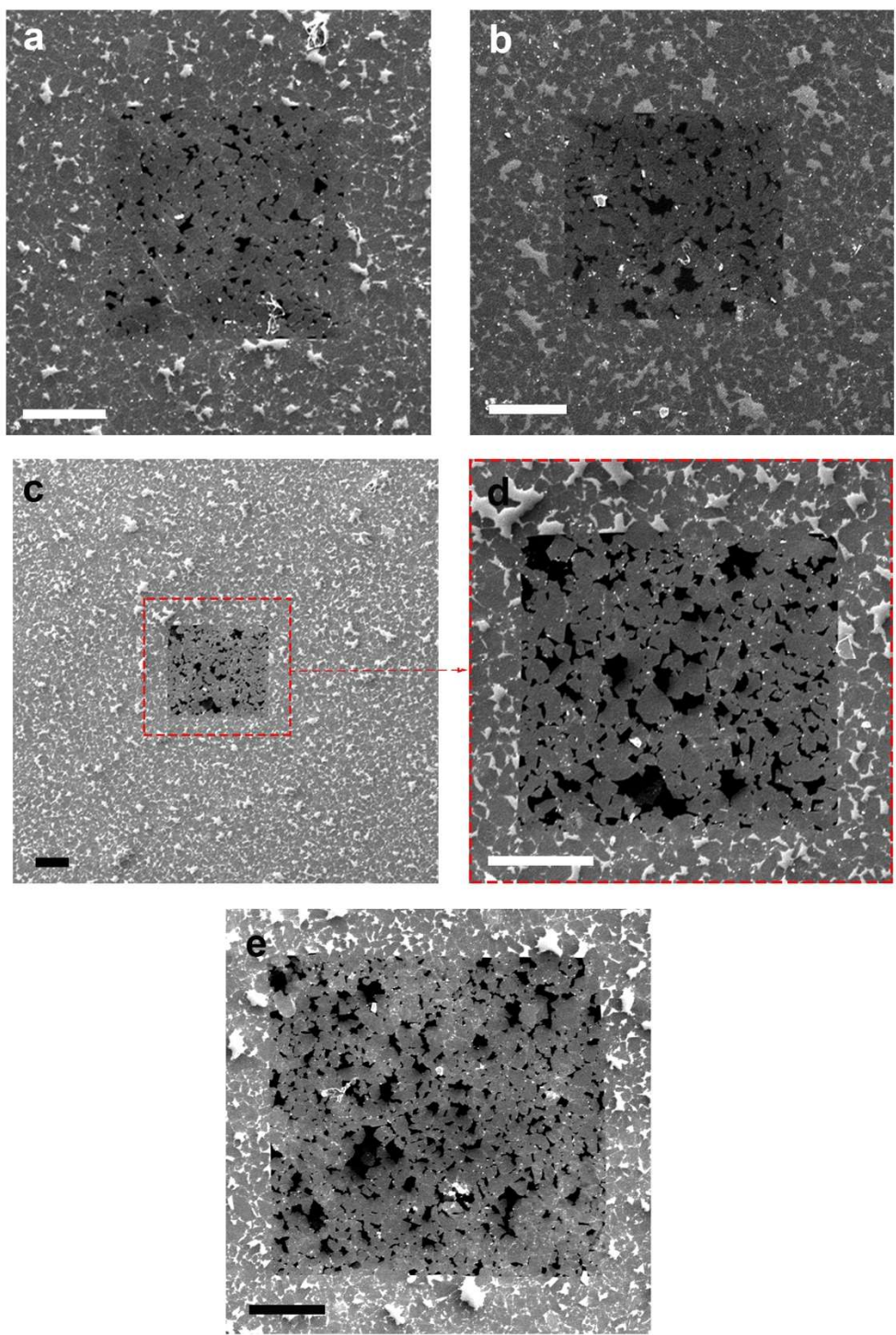

**Figure S6.** (a-e) SEM images of the transferred $Ti_3C_2T_x$ monolayer films on Si chips. The black area shows the free-standing 50-nm-thick SiNx membranes. (All scale bars = 20 μm)

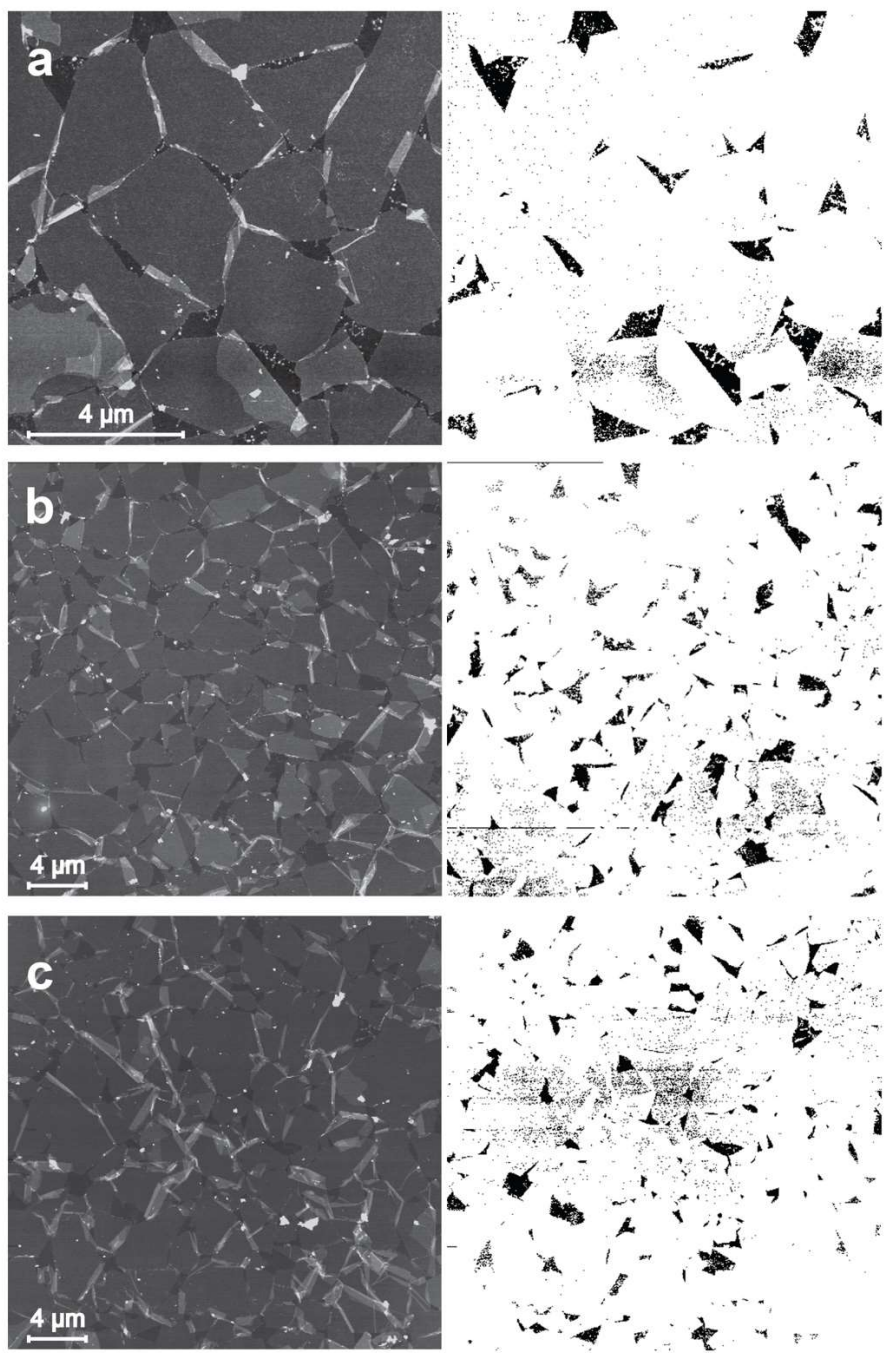

**Figure S7.** (a-c) AFM images of the transferred monolayer $Ti_3C_2T_x$ films on $SiN_x$ substrates with their corresponding analyzed images (on the right) using ImageJ software. The analyzed images show substrate coverage of (a) 93.8% (b) 93.9% and (c) 93.4 % by the film, respectively.

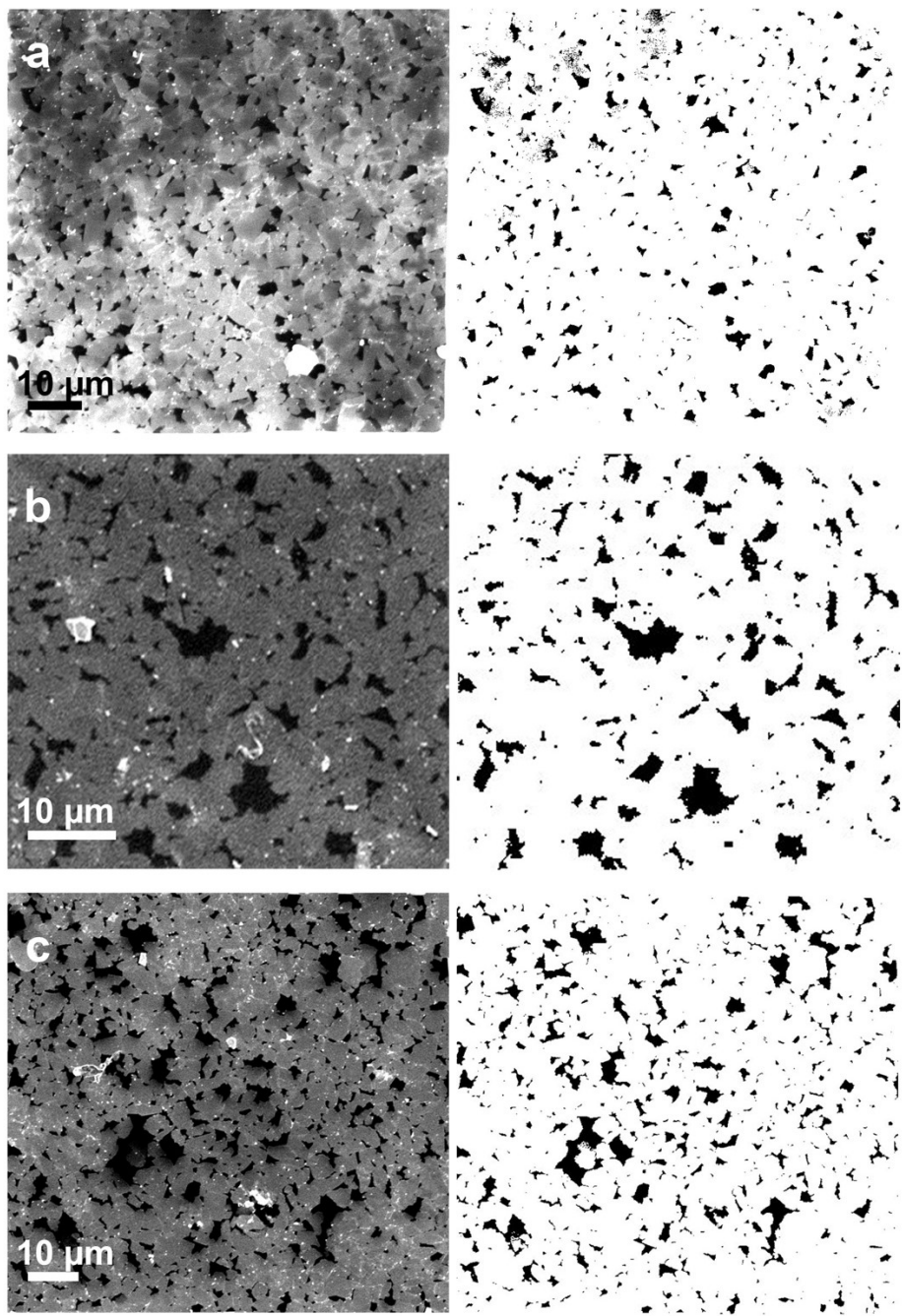

**Figure S8.** (a-c) SEM images of the transferred monolayer $Ti_3C_2T_x$ films on 50-nm-thick $SiN_x$ membranes with their corresponding analyzed images (on the right) using ImageJ software. The analyzed images show substrate coverage of (a) 94.5% (b) 89.8% and (c) 90.1% by the film, respectively.

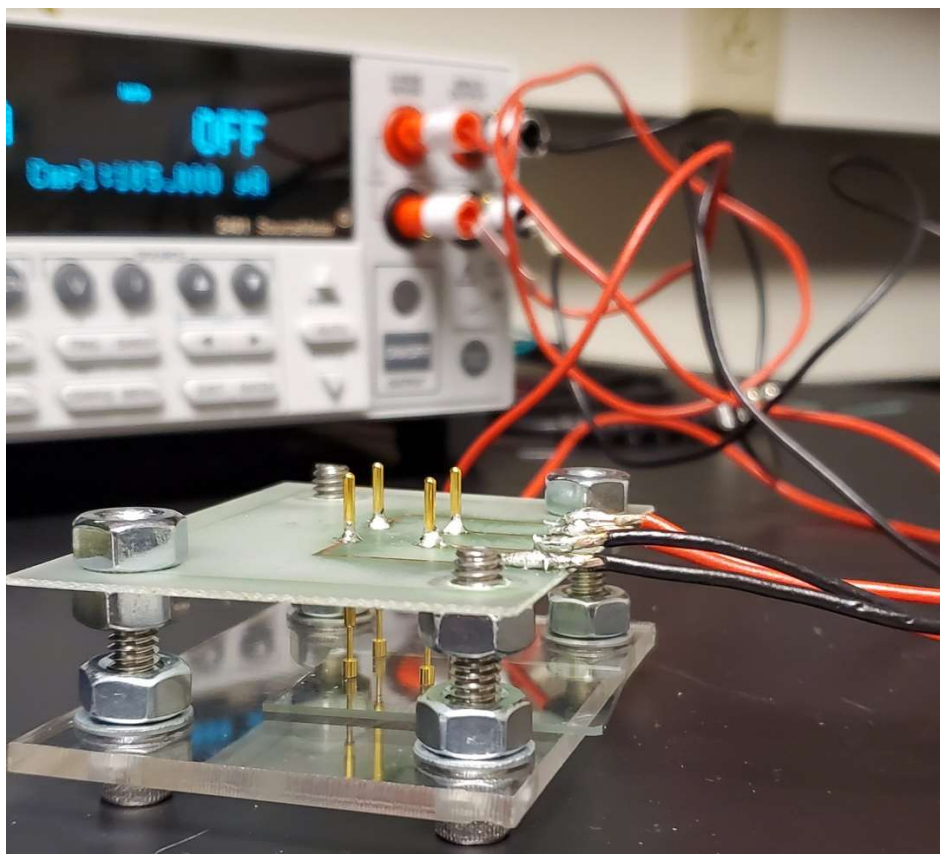

**Figure S9.** VdP experimental set-up for measurement of Ti$_3$C$_2$T$_x$ sheet resistance.